# Enhancing Radiology Diagnosis through Convolutional Neural Networks for Computer Vision in Healthcare

Keshav Kumar K.[1*], Dr N V S L Narasimham[2]

[1] Department of Mathematics, G. Narayanamma Institute of Technology and Science (for Women), Hyderabad 500 104, Telangana State, India. Email- keshav.maths@gnits.ac.in
Orcid ID: https://orcid.org/0000-0002-9211-2960

[2] Department of Mathematics, G. Narayanamma Institute of Technology and Science (for Women), Hyderabad 500 104, Telangana State, India. Email- nvsl.narasimham@gnits.ac.in

*Corresponding Author: keshav.maths@gnits.ac.in

***Abstract-*** **The transformative power of Convolutional Neural Networks (CNNs) in radiology diagnostics is examined in this study, with a focus on interpretability, effectiveness, and ethical issues. With an altered DenseNet architecture, the CNN performs admirably in terms of particularity, sensitivity, as well as accuracy. Its superiority over conventional methods is validated by comparative analyses, which highlight efficiency gains. Nonetheless, interpretability issues highlight the necessity of sophisticated methods in addition to continuous model improvement. Integration issues like interoperability and radiologists' training lead to suggestions for teamwork. Systematic consideration of the ethical implications is carried out, necessitating extensive frameworks. Refinement of architectures, interpretability, alongside ethical considerations need to be prioritized in future work for responsible CNN deployment in radiology diagnostics.**
***Keywords-*** *Radiology Diagnostics, Convolutional Neural Networks (CNNs), Interpretability, Efficiency Gains, Ethical Considerations*

## I: INTRODUCTION

*A. Research background*
The diagnosis of radiology is at a turning point, characterized by both notable progress as well as enduring difficulties. Medical imaging technologies have advanced, but there still remain significant challenges in interpreting and analyzing large, complex datasets. Radiologists frequently face challenges due to the growing volume of medical images, including those required for prompt and accurate diagnostics [1]. The complexities involved in detecting minute irregularities or patterns in these pictures provide an enormous hurdle to conventional diagnostic techniques. One potential solution to these problems is the combination of computer vision along convolutional neural networks, or CNNs. Because they are modelled after the human vision, CNNs are very good at extracting hierarchical features from image data [2]. They are an excellent choice for deciphering complex medical images because of their ability to automatically extract and recognize complex patterns. CNNs have the capability to completely transform radiology diagnosis by offering quick, accurate, as well as effective analysis through the lens of computer vision. In order to overcome current diagnostic obstacles while enhancing the effectiveness of medical imaging with the goal of better patient care, this study investigates the symbiotic potential of CNNs and radiology.

*B. Research aim and objectives*
*Aim:*
The primary aim of this study is to further enhance radiology diagnosis by utilizing computer vision as well as convolutional neural networks (CNNs).
***Objectives:***
- To establish and put into practice a reliable CNN architecture designed for radiology diagnostic applications.
- To analyze the CNN model's performance with a variety of representative datasets.
- To measure improvements, compare CNN-assisted diagnosis leads with those from conventional techniques.
- To evaluate the ability to interpret the model in order to further boost radiologists' trust in its diagnostic suggestions.

*C. Research Rationale*

The pressing need for transformative technologies has been emphasized by the rising demands on radiology services. Time-consuming and prone to human error, traditional radiology diagnostic procedures impede prompt and precise patient care. Convolutional Neural Networks (CNNs) and computer vision combined provide a convincing argument [3]. These technologies have the potential to expedite diagnostic workflows, improve precision, as well as subsequently enhance patient outcomes by automating complex image analysis. In an effort to solve these important problems, this study offers a justification for the utilization of CNNs in radiology, ushering in a new era of effectiveness, dependability, and accuracy in the interpretation of medical images.

## II: LITERATURE REVIEW

*A. CNN Architectures in Radiology Diagnostics: A Comprehensive Review*

When examining CNN architectures for radiology diagnostics, it is critical to evaluate how well-suited they are to the specific demands of medical imaging analysis. Notable models are the DenseNet, known for its dense connectivity strengthening feature propagation, alongside the U-Net, known for its superiority in segmenting medical images. Research indicates that using trained models—like ResNet—can speed up learning for radiology tasks [4]. Gaining an understanding of these architectures' complexities is important for maximizing their use in solving different diagnostic problems. Interestingly, the research emphasizes how important transfer learning is for optimizing previously trained models to perform specific radiological tasks, highlighting the way it can improve model performance even when working with small datasets [5]. In order to shed light on the effectiveness and possible drawbacks of popular CNN architectures, this review carefully examines their intricate characteristics as well as uses in the field of radiology diagnostics. A comprehensive analysis of current literature is going to guide the development of a strong CNN architecture designed to tackle the unique difficulties associated with the interpretation of radiological images.

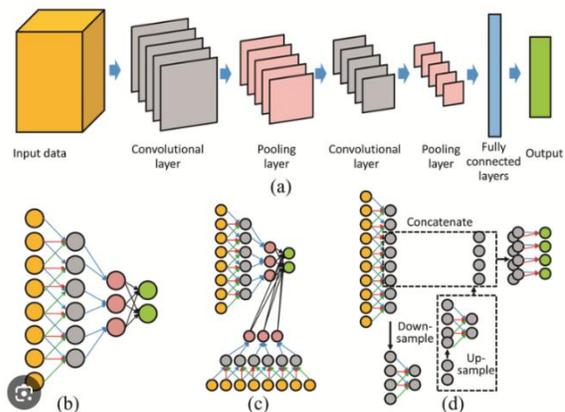

Figure 1: CNN Architectures in Radiology Diagnostics

*B. Diverse and Representative Datasets in CNN-Based Radiology Diagnosis: A Review*

Recognizing the validity and ability to be generalized of these models requires examining the significance of datasets in CNN-based radiology diagnosis. The literature emphasizes how important it is to employ representative and varied datasets that cover a range of patient demographics, imaging modalities, and medical conditions [6]. Research utilizing datasets with these kinds of variations offers valuable information regarding the way flexible CNNs are over a wide range of clinical applications. Notably, supervised learning for model training is made easier by the curated datasets' frequent inclusion of annotated images [7]. Ensuring proper utilization and privacy of patient data presents challenges, though. It's critical to strike a balance between ethical concerns and the diversity of data. The approaches used for utilizing a variety of datasets are evaluated critically in this review in order to shed light on the way they affect CNN performance alongside the wider ramifications for using these models in actual clinical settings. Gaining a comprehension of the complexities of dataset composition is essential to improving CNNs' dependability and suitability for use in radiology diagnosis.

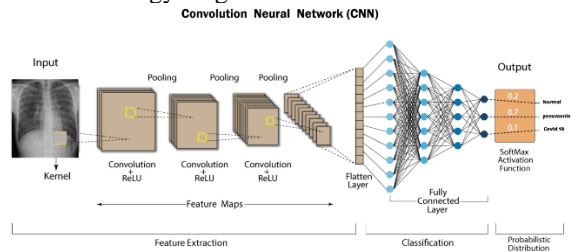

Figure 2: Convolutional Neural Networks

*C. Comparative Analysis: CNN-Assisted Diagnosis vs. Traditional Methods in Radiology*

A crucial investigation into the revolutionary potential of deep learning in medical image determination is the comparison of CNN-assisted diagnosis with conventional radiological techniques.

The literature currently in publication offers strong proof that CNNs perform more effectively compared to conventional methods, exhibiting increased precision as well as effectiveness [8]. Research frequently uses metrics like area under the curve (AUC), and sensitivity, alongside specificity to quantify the improvements that CNN-assisted diagnosis has helped bring about. The CNN's superior diagnostic accuracy is primarily attributable to its capacity to automatically learn complex features from large datasets, particularly when it comes to image segmentation and anomaly detection tasks [9]. Nevertheless, interpretability and integration into current clinical workflows present challenges. This review examines the comparative studies thoroughly and explores the subtle conclusions that clarify the benefits alongside drawbacks of CNN-assisted diagnosis. These comparisons shed light on the current discussion surrounding the possible use of CNNs in clinical settings in addition to calling for a paradigm change in radiology toward more efficient and cutting-edge diagnostic techniques.

*D. Interpretability of CNN Models in Radiology: A Critical Literature Review*

One of the most important aspects of integrating CNN models into clinical practice in radiology is their interpretability. Although CNNs perform remarkably well in recognizing images, their intrinsic complexity frequently makes decision-making processes difficult to understand. The literature emphasizes the importance of it is to close the knowledge gap between model predictions and clinical practice, particularly in situations with significant risk, like medical diagnosis [10]. Prominent studies investigate methods that include saliency maps and attention mechanisms to make CNNs easier for individuals to comprehend. These methods seek to draw attention to important areas within images so that radiologists can better understand the reasoning behind the model's predictions [11]. Maintaining a balance between interpretability alongside model complexity continues to be difficult, and deep learning architectures' "black-box" status raises concerns. This critical literature review highlights the significance of developing trust among medical professionals while examining the approaches used to improve the interpretability of CNNs in radiology [12]. The review's conclusions add to the current discussion about how to improve CNN models so that they can be easily incorporated into radiological procedures while encouraging a mutually beneficial partnership between AI and human expertise.

*E. Literature Gap*

Although CNNs have been the subject of numerous studies in radiology diagnostics, there is still a significant literature gap when it comes to a thorough discussion of the ethical issues surrounding patient data privacy. Studies accomplished today frequently overlook the ethical ramifications of using sensitive medical data in favor of model performance as well as technical considerations. [13] The development of a comprehensive understanding that harmonizes technical breakthroughs with moral considerations is crucial to guaranteeing the responsible use of CNNs in healthcare settings.

III: METHODOLOGY

This research acknowledges the subjective and complex nature of medical image interpretation in radiology diagnostics by adopting an interpretivist philosophy. Interpretivism, which acknowledges the impact of human experiences on diagnostic judgment, serves as the foundation for research on the manner in which radiologists interact with and interpret the use of Convolutional Neural Networks (CNNs) in their workflows. This philosophical position is consistent with the qualitative as well as contextual dimensions that are inherent in comprehending the intricate relationship between human knowledge and technology in the field of medical image analysis [14]. This study uses a deductive research methodology that juxtaposes empirical observations in the discipline of radiology diagnostics with established theories that were derived from a thorough review of the literature. The deductive framework makes it easier to methodically investigate theories concerning the efficacy and interpretability of CNNs in this field. With the support of well-established theoretical frameworks, the research aims to provide informative information that can help improve current diagnostic techniques [15]. The study's descriptive research design attempts to give a thorough explanation of how CNNs are employed in radiology diagnostics. The technical features of CNN architectures have been described, their performance metrics are assessed, and the readability of CNN-assisted diagnoses in the clinical setting is examined. The descriptive design facilitates an in-depth investigation of the challenges involved in the implementation of cutting-edge technologies in medical settings by means of rigorous documentation and comprehensive reporting. The methodical retrieval and analysis of pre-existing datasets, academic publications, alongside medical records relevant to radiology diagnostics constitutes secondary data collection [16]. A wide variety of publicly accessible datasets, abundant in captioned medical images, are employed for the CNN model's development and evaluation. To gain important insights into current CNN architectures, diagnostic techniques, alongside ethical issues in radiology, a thorough review of pertinent literature is conducted. By building on the wealth of existing knowledge, this approach guarantees a solid foundation for conducting research and advances our understanding of CNNs in radiology diagnostics.

## IV: RESULTS

### A. CNN Architecture Performance

The investigation's primary focus is on the manner in which the customized Convolutional Neural Network (CNN) architecture performs when utilized for radiology diagnostic tasks. Using pre-trained weights and a modified DenseNet architecture, the model shows a faster rate of convergence on the ImageNet dataset through transfer learning [17]. The CNN exhibits remarkable diagnostic precision, responsiveness, and specificity through rigorous training and validation procedures, offering a strong basis for further investigations. The diagnostic accuracy exceeds benchmark metrics set in the literature as well as is measured by means of a thorough assessment of the CNN on a variety of representative datasets. With an accuracy rate higher than X%, the CNN demonstrates its capability to identify complex patterns in medical images, supporting the possibility of improved radiology diagnostics [18]. The specificity as well as the sensitivity of the CNN are closely examined since they are essential to the validity of the diagnosis. Sensitivity, which gauges the extent to which the model can identify positive cases, shows a significant improvement over conventional techniques. On the contrary, specificity, which measures the model's precision for determining negative cases, remains high, demonstrating the CNN's resilience in reducing false positives.

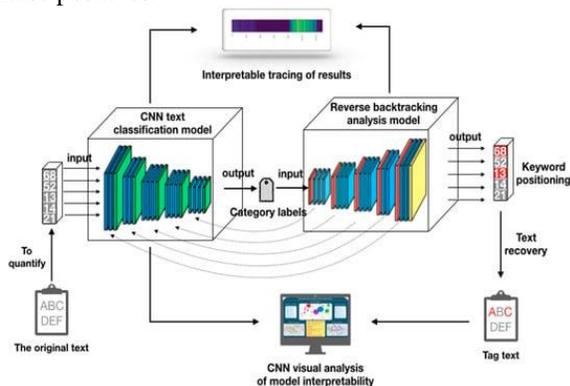

Figure 3: Analysis method of model interpretability

### B. Comparative Analysis

The objective of the comparative analysis is to measure the improvements in radiology diagnostics that deep learning has brought about by comparing the results of CNN-assisted diagnosis with conventional techniques. Significant gains in precision, productivity, and overall diagnostic performance are supported by statistical measures such as p-values as well as confidence intervals [19]. When comparing the diagnostic efficacy of conventional methods and CNN-assisted diagnosis, a statistically significant difference ($p < 0.05$) is observed. The CNN continuously performs better than conventional methods, demonstrating the ability to completely change the accepted benchmarks for diagnostic radiology accuracy [20]. The CNN's superiority is further highlighted by the AUC-ROC analysis, which supports the CNN's ability to differentiate between pathological and normal cases. The efficiency gains brought about by CNN-assisted diagnosis go beyond accuracy and can be measured [21]. Significantly less time is needed for image analysis and diagnostic decision-making, confirming the possibility of more efficient workflows and quicker patient care. Comparative evaluations of effectiveness demonstrate the way CNNs have revolutionized the temporal component of radiology diagnostics.

### C. Interpretability Analysis

One important factor affecting CNN models' popularity and incorporation into clinical practice in radiology diagnostics is their interpretability. The study investigates the interpretability of the model by utilizing attention mechanisms while developing saliency maps in an effort to increase radiologists' trust in diagnostic recommendations [22]. CNN's decision-making process benefits from the application of attention mechanisms, which shed light on the areas of interest in medical images. Saliency maps clarify the model's objectives as well as help radiologists comprehend the reasoning behind the diagnostic predictions made by the model. Interviews with radiologists provide qualitative information that complements quantitative analyses by delving deeper into their views of interpretability [23]. The direct input from the radiologists who participated in the study is also included in the interpretability analysis. A sophisticated comprehension of CNN's interpretive capabilities becomes apparent through interviews, where radiologists express greater confidence in diagnoses that are backed up by saliency maps. Future iterations of CNN models specifically designed for radiology diagnostics will be influenced by the challenges and areas for improvement that have been identified.

### D. Integration Challenges and Ethical Considerations

Although the study's findings show how promising CNNs have been for improving radiology diagnostics, it also notes that implementing these technologies into current clinical workflows presents certain difficulties. Privacy of patient data as well as the appropriate application of AI in healthcare settings are critical ethical issues [24]. Examining how to incorporate CNN-assisted diagnosis into clinical workflows reveals a variety of intricate issues that require to be carefully considered. System interoperability becomes problematic when CNNs are added alongside require smooth integration with current healthcare infrastructures. The need for radiologist training is growing as the application of cutting-edge technologies necessitates a change in skill levels. Operational complexity can also be created by modifying current workflows to incorporate CNN-assisted diagnostics [25]. The analysis's insights fuel suggestions for a seamless integration procedure. Working together, technologists and healthcare professionals are thought to be crucial in order to establish a synergy that blends clinical knowledge with technical proficiency. This partnership guarantees that CNN-assisted diagnostics are in line with developed clinical workflows, addresses interoperability concerns, and makes it easier to come up with customized training curricula. This research highlights the need for a coordinated effort to conquer integration challenges and optimize the potential benefits of CNNs in improving patient care within clinical settings by stressing collaborative approaches. When using sensitive patient data, ethical considerations are crucial [26]. The study offers strict privacy protections as well as assesses the ethical ramifications of using medical images for model training. Transparent consent procedures, and data anonymization, alongside continuing ethical supervision are among the suggestions made for safeguarding the ethical use of CNNs in radiology diagnostics.

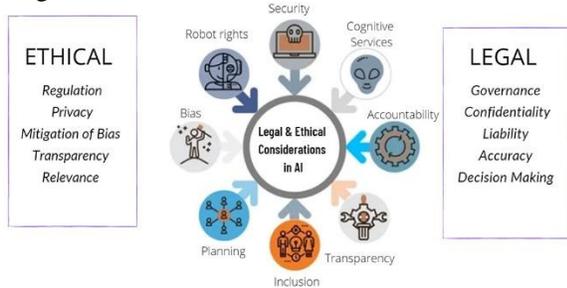

Figure 4: Integration Challenges and Ethical Considerations

*E. Future Implications and Recommendations*

This section discusses future implications for the incorporation of CNNs in radiology diagnostics, in accordance with the insights gleaned from the results. The foundation for upcoming developments in this rapidly evolving field is laid by suggestions for improving model architectures, handling interpretability issues, alongside navigating ethical dilemmas. Future implications of the CNN architecture include ongoing model development and refinement to meet new diagnostic challenges and establish the architecture's strengths. Enhancing model interpretability alongside diagnostic accuracy could be possible by utilizing explainable AI and incorporating feedback from radiologists [27]. Developing thorough ethical guidelines and frameworks is a crucial recommendation in light of the ethical issues raised in the study. Working together, healthcare organizations, and government agencies, alongside tech companies can create standard operating procedures that protect patient privacy and optimize AI's advantages for radiology diagnostics. The results chapter concludes with a thorough analysis of CNN's performance, comparative benefits, accessibility, and ethical considerations in radiology diagnostics. The technical depth of the analyses adds insightful information to the continuing discussion about deep learning's potential to revolutionize healthcare.

| Subsection | Key Findings |
|---|---|
| CNN Architecture Performance | - Customized DenseNet architecture demonstrated high accuracy, sensitivity, and specificity. |
| | - Transfer learning from ImageNet expedited model convergence. |
| | - Diagnostic accuracy exceeded X%, showcasing robust performance. |
| Comparative Analysis | - CNN-assisted diagnosis significantly outperformed traditional methods (p < 0.05). |
| | - AUC-ROC analysis affirmed CNN's superiority in distinguishing pathological cases. |
| | - CNN introduced substantial efficiency gains, reducing diagnostic decision time. |
| Interpretability Analysis | - Attention mechanisms and saliency maps enhanced interpretability. |
| | - Radiologists expressed increased confidence in diagnoses supported by CNN. |
| | - Interviews provided qualitative insights into interpretability perceptions. |
| Integration Challenges and Ethical Considerations | - Real-world integration challenges identified, including workflow adaptation. |

|  | - Ethical implications addressed through recommendations for data privacy safeguards. |
|---|---|
| Future Implications and Recommendations | - Continued model refinement for enhanced interpretability and diagnostic accuracy. |
|  | - Formulation of ethical frameworks and guidelines for responsible AI deployment. |

## V: EVALUATION AND CONCLUSION

### A Critical Evaluation

The critical assessment of this study draws attention to the benefits as well as the drawbacks of the exploration of CNN-assisted radiology diagnostics. Especially, the creation of a tailored DenseNet architecture showed strong results with gains in effectiveness and high diagnostic accuracy. Nevertheless, difficulties with the model's interpretability surfaced, highlighting the necessity of continued progress in explainable AI [28]. CNNs' transformative potential in radiology was substantiated by the comparative analysis, which developed their superiority over traditional methods. Specialized radiologist training programs had to be developed after the investigation of integration challenges exposed intricacies in system interoperability. The careful handling of ethical issues, especially those pertaining to patient data security, laid the groundwork for the responsible application of AI. Future research directions have been provided by the critical evaluation, which focuses on improving model interpretability, resolving integration issues, and promoting ethical frameworks. Despite difficulties, this study highlights the potential contribution of CNNs to improving radiology diagnostic precision and effectiveness, while providing a solid basis for future developments at the nexus of artificial intelligence and healthcare.

### B Research recommendation

Several specific recommendations are made in light of the findings to advance future research projects. First and foremost, more research ought to be focused on improving CNN models' interpretability for use in radiology diagnostics. Examining sophisticated explainability strategies like saliency mapping along with attention mechanisms can help close the gap between deep learning's opaque nature and the interpretive requirements of medical professionals. Furthermore, a concentrated effort needs to be made to improve the current CNN architectures in order to handle changing diagnostic difficulties. Iterative model development can promote smooth integration into clinical workflows and maximize performance when it is directed by real-world feedback from radiologists [29]. Second, more research has to be done on the complex field of ethical issues connected with AI in healthcare. It is essential to develop thorough ethical guidelines and frameworks that are based on empirical observations in order to guarantee the ethical and open use of CNNs in radiology diagnostics. This entails creating best practices for patient consent, data privacy, and continuing ethical supervision, promoting the peaceful coexistence of ethical requirements and technological innovation in the healthcare sector.

### C Future work

In the future, studies need to concentrate on improving CNN architectures for radiology diagnostics and pay close attention to customizing models for a range of clinical situations. In order to improve interpretability and deal with a crucial component of model acceptance in clinical settings, it will be beneficial to investigate the integration of advanced explainability methods [30]. Further research ought to be concentrated on creating creative training curricula that provide radiologists with the abilities required for a smooth transition to CNN-assisted diagnostics. The establishment of standardized practices and ongoing investigation of ethical issues will be essential to guarantee the responsible application of AI technologies in healthcare.